\documentstyle[11pt,newpasp,epsf,twoside]{article}
\def\edcomment#1{\iffalse\marginpar{\raggedright\sl#1\/}\else\relax\fi}
\reversemarginpar

\begin{document}
\title{Triggering star formation by galaxy-galaxy interactions}
 \author{Tissera, P. B. }
\affil{Institute of Astronomy and Space Science, Conicet}
\author{Alonso, M. S.}
\affil{Complejo Astron\'omico El Leoncito,Conicet}
\author{ Lambas, D. G. \& Coldwell, G.}
\affil{IATE, Observatorio Astron\'omico C\'ordoba. Conicet}

\begin{abstract}
We analyzed the effects of having a close companion on the
star formation activity of galaxies in 8K galaxy pair catalog 
selected from the 2dFGRS. We found that, statistically,
galaxies with $r_{\rm p} <  25 {\rm h^{-1}}$ kpc and 
$\Delta V <  100 {\rm km/s}$
 have enhanced star formation 
with respect to isolated galaxies with the same luminosity
and redshift distribution. 
Our results suggest that the physical processes at work during tidal
interactions can 
 overcome the effects of environment, expect in  dense regions. 
\end{abstract}

%\section{Analysis and Discussion}

Different physical processes  are thought to be involved in
the  regulation of the star formation in galaxies, but it is now accepted that
mergers and interactions play an important role as different
observational 
 and numerical
works have shown.
However, only recently it has been possible to study the effects of interactions
on statistical basis. Barton et al. (2000) found that the star formation
activity correlates with proximity in projected  distance and velocity
difference by analyzing a sample of approximately 200 pairs.
The release of  the 2dFGRS  
 opened the possibility of carrying out an analysis of
star formation in galaxy pairs on statistical basis. By cross-correlating
the galaxy pair catalog with the group catalog of  
 Merch\'an \& Zandivarez (2003, in preparation)  
 galaxy pairs  where classified according to the environment.
 In this work
we focuse on 
 the question of how close have to be galaxies in order to
show  star formation activity enhancement.
For this purpose, we will always compare the properties of galaxies in pairs
with those of isolated galaxies selected from the 2dF with the same
redshift and luminosity distribution.

%\section{Analysis and Results}

We estimated the birth rate parameter $b= SFR/<SFR>$ for galaxies in the 
2dFGRS from the correlation between the  $\eta$ spectral type
and the $H_{\alpha}$ equivalent width
 (see Lambas et al. 2003 for details).
Note that the SFR are deduced from $H_{\alpha}$ and no dust effects have been 
considered.
We calculated the  mean $b$ parameter for the 
 neighbors within concentric spheres centered
at a given galaxy. The center galaxies were chosen according to their
spectral parameters $\eta$ (see Fig1a)
%: high star formation centres $\eta > 3.5$,
%late-type galaxies $\eta > -1.4$ and  any $\eta$ value.
%We found that  as further away neighbors are cast out, the star formation
%activity increases (Fig.1a). 
A similar calculation was performed by binning
in relative velocity separation. From this analysis we found that galaxies
in pairs satisfying 
$r_{\rm p} < 100 {\rm h^{-1}}$ kpc 
 and $\Delta V  < 350 {\rm km /s}$ is enhanced.
By applying these two criteria, 
we selected approximately 9000 pairs with $ z < 0.1$ in high
and low density environments.
We also constructed a galaxy control sample by identifying galaxies
with non close companion
in the field (or cluster) with the same redshift and luminosity distribution as galaxies in pairs.
We then estimated the mean $b$ parameters for galaxies in pairs in projected and
relative velocity bins, and the mean $b$ value of the control sample.
By equating them, we found that the star formation activity is statistically enhanced
by the presence of a companion from $r_{\rm p} \approx 25  {\rm h^{-1} kpc}$
and $\Delta V  < 100 {\rm km /s}$
 in comparison to isolated galaxies (see Fig.1b).
%In the small box we plotted the fraction that galaxies in pairs with $b > \bar{b}$
%finding similar results.
This analysis extended to galaxy pairs in groups (Alonso et al. 2004)
yields 
similar results, except for galaxy pairs in very high density regions where
the both colors and $b$ show no star formation activity in the present.
 Hence, statistically, tidal fields
seem to be efficient in enhancing the star formation activity for very
close pairs. The relative velocity and projected separation
thresholds are  independent of environment, suggesting
that galaxy-galaxy interactions can be a main, ubiquitous motor of star
formation activity in the Universe.

\begin{figure}[h]
\plottwo{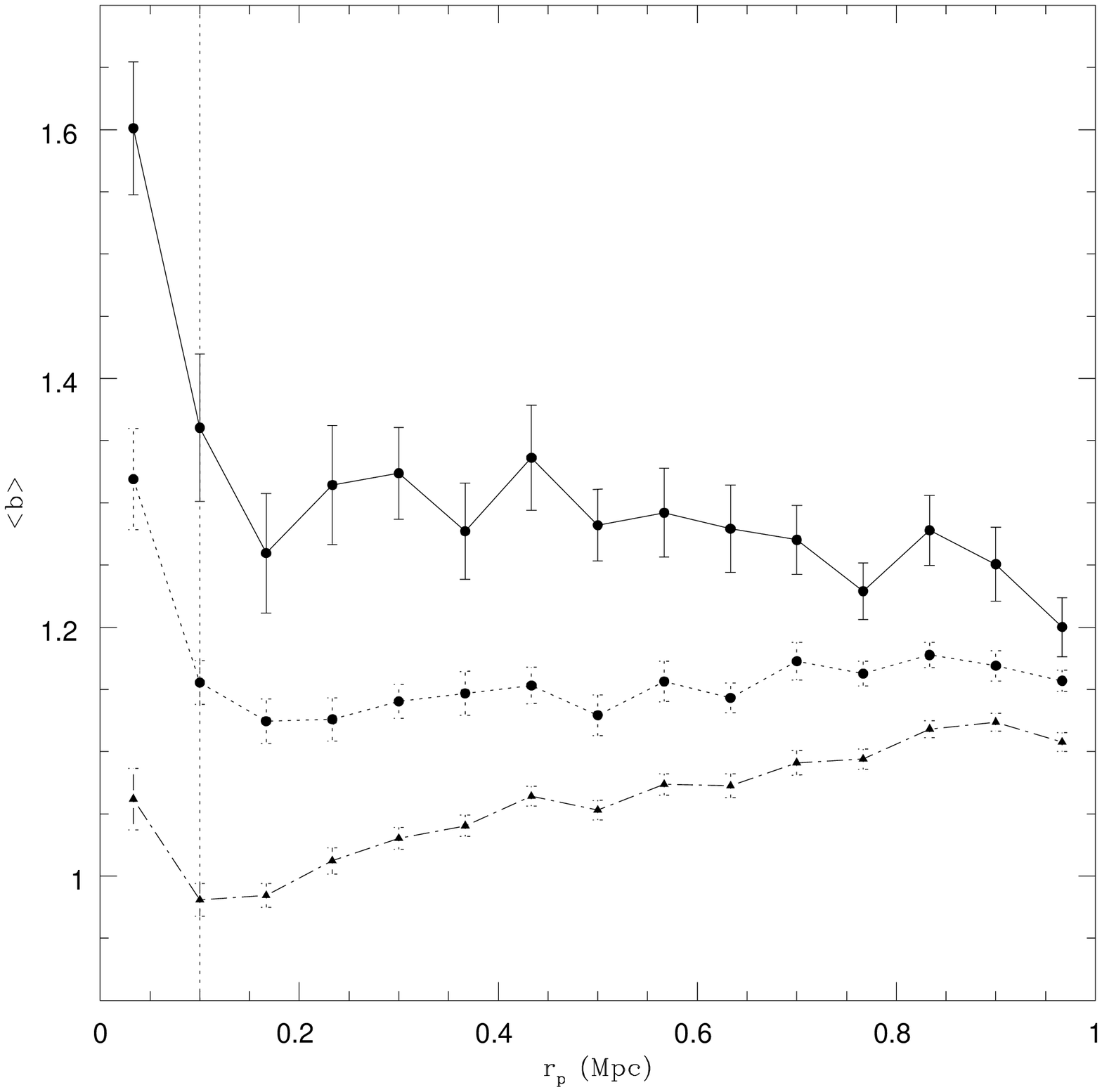}{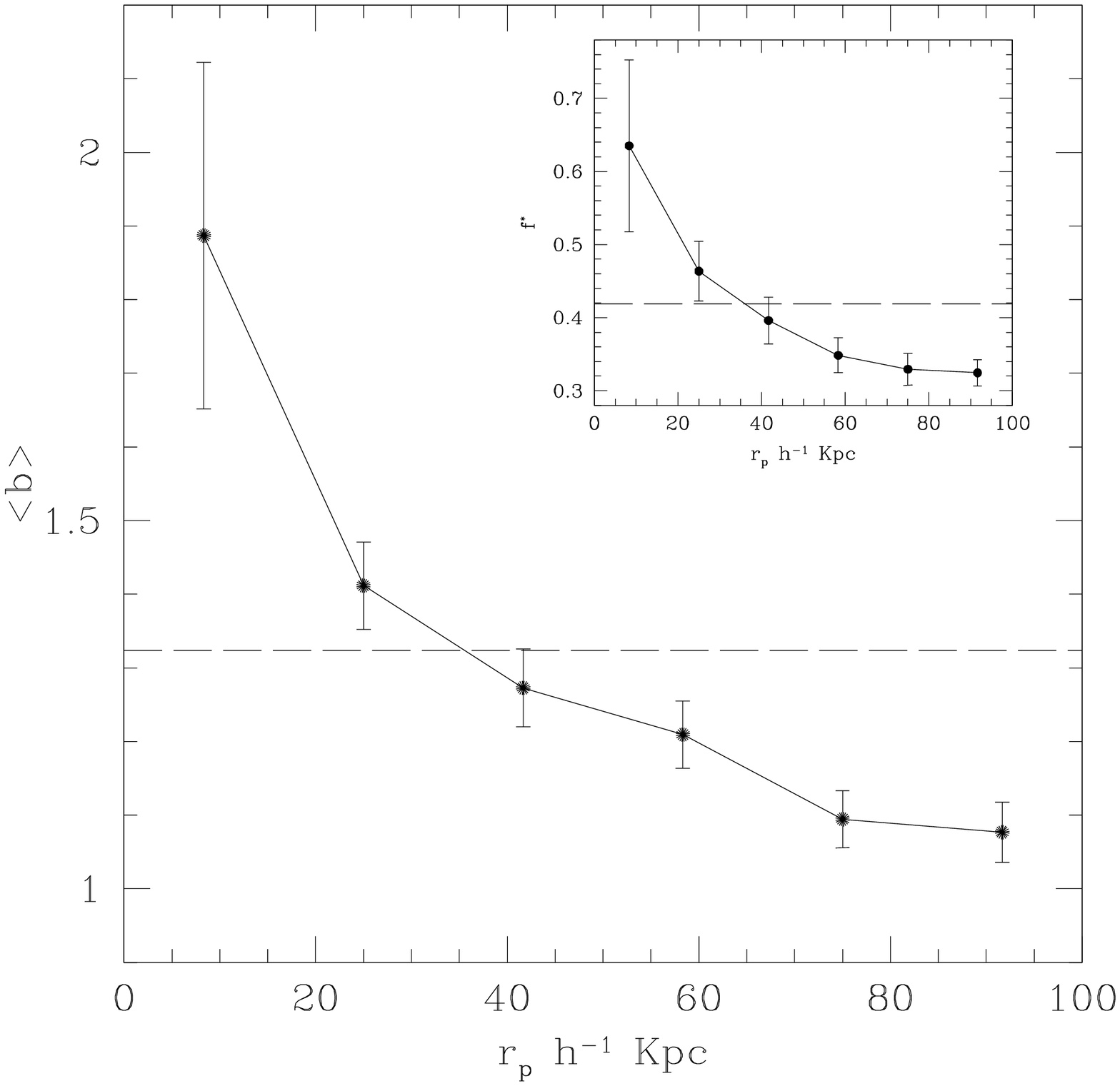}
\caption{a)Mean birthrate parameter $b$ as a function of relative
projected separation $r_p$
of galaxies with $\eta >3.5$ (solid line),
$\eta >-1.4$ (dotted line) and  no $\eta$ restriction (dotted-dashed line).
The dotted vertical line depict the spatial separation threshold
identification. b) Mean $b$ parameters estimated in projected distance
bins for galaxies in interacting pairs in the field.
The small box shows  the fraction $f^\star$ of galaxies
with $b>\bar{b}$.
The dashed horizontal lines
represent the mean $b$ parameter for the corresponding  control sample.
}
\end{figure}

%\acknowledgments{PBT thanks the SOC and IAU for making this presentnation
%possible. This work was partially supported by Conicet, Secyt and Fundaci\'on 
%Antorchas}


\begin{references}
\reference Alonso, M. S,  Tissera, P. B.,  Coldwell, G., Lambas, D. G., 2003,
submitted.
\reference  Barton E. J., Geller M. J., Kenyon S. J., 2000, ApJ, 530, 660  
%\reference Merch\'an, M., Zandivarez, A., 2002, MNRAS, 335, 216
%\reference Madgwick D. S. et al., 2002a, MNRAS, 333, 133 
\reference Lambas, D. G., Tissera, P. B., Alonso, M. S. Coldwell, G. 2003,
MNRAS in press 
\end{references}
\end{document}